\documentstyle[11pt,newpasp,twoside,epsf]{article}
\def\edcomment#1{\iffalse\marginpar{\raggedright\sl#1\/}\else\relax\fi}
\reversemarginpar

\begin{document}
\title{High Frequency Peakers: Baby Radio Sources?}
 \author{Sara Tinti}
\affil{SISSA/ISAS, Via Beirut 4, Trieste, Italy, 34014}
\author{Daniele Dallacasa}
\affil{ Dipartimento di Astronomia, via Ranzani 1, Bologna, Italy, 40127}
\author{Gianfranco De Zotti}
\affil{INAF, Osservatorio Astronomico, Vicolo dell'Osservatorio 5, Padova, Italy, 35122}
\author{Carlo Stanghellini}
\affil{Istituto di Radioastronomia-CNR, C.P. 161, Noto, Italy, 96017}
\author{Annalisa Celotti}
\affil{SISSA/ISAS, Via Beirut 4, Trieste, Italy, 34014}
\begin{abstract}
We present the results of second epoch simultaneous multifrequency observations
of 45 sources out of the 55 in the sample of High Frequency Peakers (HFP) 
by Dallacasa et al. (2000). Seven of the stellar sources 
(i.e. 16\% of the total number of sources and 26\% of the stellar fraction) 
 no longer show a convex spectrum and are thus likely
to be flat spectrum quasars in an outburst phase during the first epoch of 
observations. The rest frame peak frequency distribution of HFP quasars
extends up to $\sim 45$ GHz while that of HFP galaxies is confined to $\le 17$
GHz. Imaging at 1.465 GHz and 1.665 GHz has revealed extended emission for 31\%
 of sources, a substantially larger fraction than found for GPS sources 
($\sim 10\%$). In the case of HFP quasars, but not of HFP galaxies, extended 
emission is associated with strong variability, consistent with a core-jet
 structure.    
 
\end{abstract}

\section{Scientific background}
The Compact Steep Spectrum (CSS) and Gigahertz Peaked Spectrum sources (GPS)
are two classes of intrinsically compact objects (linear size $<$ 1 kpc) 
defined on the basis of their spectral properties: the overall shape is convex
with the position of the turnover between 100 MHz (CSS) and few GHz (GPS) 
and the spectral index at high frequencies is steep (see O'Dea 1998). 

The currently favoured model (the {\textit{youth scenario}}) relates the small 
linear size of GPS/CSS sources to their age, implying that they are 
the progenitors of extended radio sources. Both 
the spectral analysis 
(Murgia et al. 1999) and the dynamical studies of the separation 
speed of the hot spots (Owsianik \& Conway 1998; Owsianik, Conway, \& Polatidis
 1998) provide strong evidence in favour of this hypothesis indicating  
age values of $10^2-10^3$ years. 

Observational studies of the population of GPS/CSS sources have led to 
the discovery  of a correlation between the radio turnover frequency 
($\nu_{max}$) and the projected angular size ($\theta$) (O'Dea \& Baum 
1997; Fanti et al. 2002). This relationship is expected from  
synchrotron self absorption, where
$\theta^2\propto \nu_{max}^{-5/2}$ , although free-free absorption
could also play a role (Bicknell, Dopita, \& O'Dea 1997)). In the youth 
scenario this means that the youngest objects have the highest turnover 
frequencies.

Objects with turnover frequencies 
above 5 GHz, which is the limit of GPS samples, would represent 
smaller/\textit{younger} sources. We call them High Frequency 
Peakers (HFP). One of the disadvantages of this approach is the contamination 
by beamed radio sources like blazars which possess radio spectra peaking 
above a few GHz as the result of self-absorbed synchrotron emission from the 
jet base.

Two epochs of simultaneous multifrequency VLA observations have been 
carried out to investigate the reliability of the selection criterium based
 on the spectral properties.
Here we present the results of the second epoch of observations and we 
compare them with those of the first epoch (see Tinti et al., in preparation).

\section{The Bright HFP Sample}
The complete sample of candidate bright HFPs has been obtained by
Dallacasa et al. (2000). Sources with 
$S_{4.9 GHz} \geq$ 300 mJy and inverted spectra (slope steeper than -0.5,
 S$\propto\nu^{-\alpha}$) were selected in the area of the 87GB catalogue 
(4.9 GHz)  by cross correlating it with the NVSS catalogue (1.4 GHz). 
The sample was then ``cleaned'' from variable flat spectrum sources by
means of simultaneous multifrequency VLA observations at 1.365, 1.665, 4.535, 
4.985, 8.085, 8.485, 14.96 and 22.46 GHz. 
The resulting sample consist of 55 sources with radio spectra peaking at
observed frequencies ranging from a few GHz to 22 GHz.

The selection criterium is independent of optical identification and 
redshift. As a result both galaxies and quasars are present in the sample,
which is composed of 20\%  galaxies, 65\%  stellar 
objects and 15\% still unidentified objects.
For comparison, galaxies and quasars are almost equally represented in 
GPS samples and most CSS sources are galaxies.    
This confirms that the fraction of galaxies, as opposed to quasars, 
decreases in samples selected at increasing turnover frequency 
(Fanti et al. 1990; O'Dea 1998; Stanghellini et al. 1998).

Additional observations in the same bands were carried out with the VLA 
for 45 sources of the original sample to study the variability in both flux 
density and spectral shape.
Moreover this subsample has been imaged at 1.465 and 
1.665 GHz by means of sensitive and longer observations in order to determine 
if there is any extended emission surrounding the compact source.

\vspace{-0.3cm}
\section{Spectral analysis}
In order to estimate the peak flux density and frequency of the HFP 
sources, we have fitted the simultaneous radio spectra at the two epochs
using an hyperbolic function. 
\begin{figure*}   
\plotfiddle{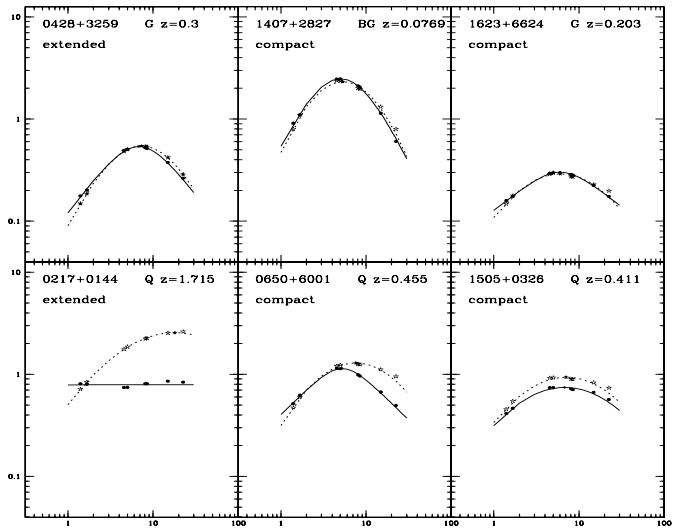}{6cm}{0}{50}{40}{-160}{-90}
\caption{Radio spectra of HFP's quasars and galaxies. Stars and filled 
circles represent respectively the first and the second
epoch of simultaneous multifrequency VLA data, while lines indicate the 
polynomial fit to the data-points.}
\end{figure*}
A considerable fraction of sources (16\%), identified as stellar objects
(i.e. 26\% of such objects), do not show a convex spectrum at the 
second epoch and are flat over the full frequency range sampled by 
our observations.
Therefore they should be classified as flat spectrum sources, 
plausibly in an outburst state in the first epoch. Among the sources which
preserve the convex spectral shape, HFP quasars are found to be more variable 
in radio flux density than galaxies (see Figure 1). 

In Figure 2, the HFPs with known redshifts are plotted together with the 
CSS sample of Fanti et al. (1990), the GPS samples of Stanghellini et al. 
(1998) and Snellen et al. (1998) in the turnover frequency-linear size plane.
The line is the expected correlation (in the source frame) in the case of 
synchrotron self absorption (Fanti et al. 2002).
The distribution of HFP sources is consistent with the correlation obeyed
by the other classes.

In GPS samples it has been found that quasars typically 
peak at higher frequencies than galaxies (Snellen et al. 1997; 
Stanghellini et al. 1998). 
Peak frequencies in the bright HFP sample are on average about a factor 5 
higher than in GPS samples but still the galaxies cover a range of values 
lower than the quasars do (Figure 2, right panel).
\begin{figure}   
\plottwo{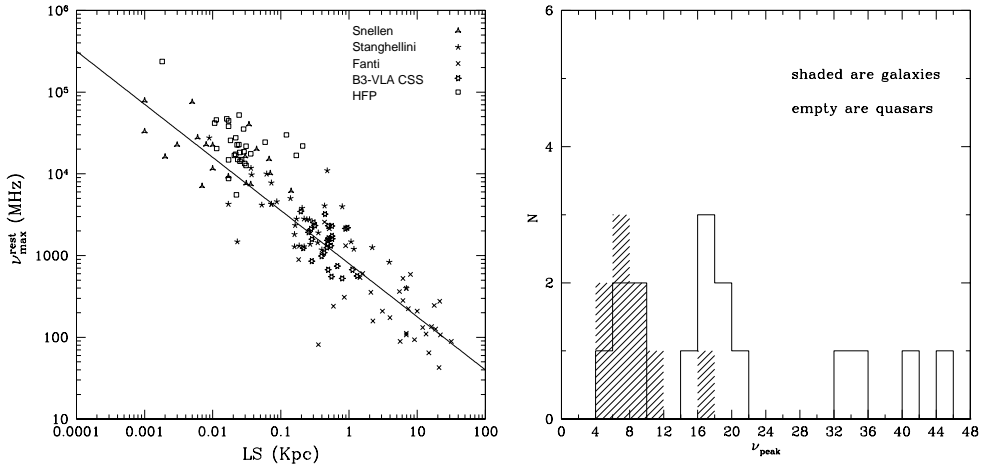}{tinti_fig3.epsi}
\caption{Rest frame turnover frequency vs. linear size for CSS/GPS samples and 
for the HFP sample(left). Rest frame turnover frequency distribution for 
galaxies and quasars at the second epoch (right).}
\end{figure}

\vspace{-0.4cm}
\section{Extended emission}
Within our sample 14 (31\%) sources show extended emission
(2 galaxies, 9 quasars and 3 objects without optical 
identification) on scales ranging from 14 to 55 arcsec.
Stanghellini et al. (1990) and Baum et al. (1990) found a fraction of 10\% 
of GPS sources displaying diffuse and faint extended emission on arcsecond 
scales near/or around the core.

The youth scenario and the presence of radio emission on scales 
of tens of kpc could be reconciled by the hypothesis of recurrency proposed 
by Baum et al. (1990): the extended emission can be the relic of a previous  
large scale radio source while the peaked spectrum is due to the young 
central component that is propagating out admist the relic of the previous 
epoch of activity.

Eight out of the nine quasars with extended emission have
 strongly variable spectra while the only two extended 
galaxies in the sample both have non variable and sharply peaked spectra.
The recurrent model seems to be suitable for radio galaxies,
while radio quasars can be more easily explained
as beamed objects whose jet is pointing towards the observer and 
whose radio emission is dominated by a single compact component.
Milli-arcsecond images at different frequencies can distinguish 
between a core-jet structure, 
typically associated to blazars, and the CSO-like structure  
expected for young or recurrent objects. 
\vspace{-0.2cm}
\section{Conclusion}
The fraction of quasars compared to galaxies is substantially higher in
our sample than in samples of GPS sources. 

Although in the literature GPS sources are frequently referred to as 
non variable (but see Stanghellini 2003), many of our sources and 
especially quasars are found to be variable. 

A substantial fraction (31\%) of our HFP sources show extended
emission which, in the case of quasars, is associated in most cases 
with strong variability. On the other hand, the two galaxies with 
extended emission are non variable. This adds to the body of evidence
that GPS galaxies and quasars 
likely represent a different phenomenon and that for quasars
the beamed emission from compact components id dominant.

Spectral selection relying on high turnover frequencies is an easier method 
for selecting candidate young sources than the morphological selection.
However the contamination of spectrally selected samples 
by Doppler boosted objects increases with turnover frequency.

We plan to complete the analysis and the identification of the sample 
using high resolution observations to study the mas morphology.
 VLBA observations, at two different frequencies in the optically 
thin part of the spectrum, have been already carried out
 and the analysis of the data is in progress.
 
\end{document}